\documentclass[journal]{IEEEtran}
\usepackage{cite}
\usepackage{multirow}
\usepackage{upgreek}
\usepackage{placeins}

\ifCLASSINFOpdf
  \usepackage[pdftex]{graphicx}
 \graphicspath{{}}
  \DeclareGraphicsExtensions{.pdf,.jpg,.png}
\fi

\usepackage{amsmath}

\ifCLASSOPTIONcompsoc
  \usepackage[caption=false,font=normalsize,labelfont=sf,textfont=sf]{subfig}
\else
  \usepackage[caption=false,font=footnotesize]{subfig}
\fi

\usepackage{stfloats}

\usepackage{url}
\usepackage{multicol}

\usepackage{xcolor}

\newcommand{\markup}[1]{#1}

\begin{document}

\title{High-efficiency refracting millimeter-wave metasurfaces}

\author{Andreas~E.~Olk, 
		Pierre~E.~M.~Macchi, 
        and David~A.~Powell,~\IEEEmembership{Senior Member,~IEEE}
\thanks{A. E. Olk  and P. E. M. Macchi are affiliated with IEE S.A., Luxembourg.}
\thanks{A. E. Olk and D. A. Powell are affiliated with the University of New South Wales, Canberra, Australia.}
\thanks{e-mail of corresponding author A. E. Olk: andreas.olk@iee.lu }

}


\maketitle

\begin{abstract}

Printed circuit metasurfaces have attracted significant attention in the microwave community for their versatile wavefront manipulation capability. Despite their promising potential in telecommunications and radar applications, few transmissive metasurfaces have been reported operating at millimeter-wave frequencies. Several secondary effects including fabrication tolerances, interlayer near-field coupling and the roughness of conductors are more severe at such high frequencies and can cause significant performance degradation. Additionally, very accurate experimental techniques are required in order to characterise these effects. %
In this work, we present highly efficient refracting metasurfaces operating at 83\,GHz. We use a
synthesis technique that minimizes performance degradation due to effects such as interlayer near-field coupling and  conductor roughness. Our experimental characterization includes an accurate determination of the intensity of all forward propagating Floquet harmonics in a broad frequency range. The experimental data shows very good agreement with full-wave simulation and verifies our synthesis method.


\end{abstract}


%
\IEEEpeerreviewmaketitle

\section{Introduction}

\IEEEPARstart{T}{he} emerging technology of metasurfaces represents a versatile concept for the manipulation of electromagnetic waves \cite{yuFlatOpticsDesigner2014,glybovskiMetasurfacesMicrowavesVisible2016,dingGradientMetasurfacesReview2018}. Due to the ease of fabrication using planar circuit manufacturing, the research community recognizes particular application potential for the microwave frequency range. Indeed, a variety of different metasurface devices have been proposed within the past decade including transmitarrays \cite{KAOUACH20161275,ryanWidebandTransmitarrayUsing2010}, flat lenses \cite{saeidiFigureMeritFocusing2015,chenDesignExperimentalVerification2019} and leaky wave antennas \cite{abdo-sanchezLeakyWaveAntennaControlled2018,liHolographicLeakywaveMetasurfaces2015}. 
Huygens metasurfaces have received significant attention as they feature near unity transmission and suppress reflection artifacts efficiently \cite{pfeifferMetamaterialHuygensSurfaces2013a,epsteinHuygensMetasurfacesEquivalence2016,jiaLowreflectionBeamRefractions2015}. For microwave frequencies, Huygens metasurfaces are usually manufactured in printed circuit board (PCB) processes with three structured copper layers separated by low loss dielectric substrates \cite{epsteinHuygensMetasurfacesEquivalence2016,jiaLowreflectionBeamRefractions2015,wong_design_2014}.

Existing works on microwave metasurfaces concentrate mainly on frequencies below 40\,GHz. However, for novel technologies in telecommunication and radar, there is an ongoing trend towards higher frequencies in the millimeter wave (mm-wave) frequency range \cite{7762824,5966314,6415978}. For 5G mobile networks and automotive radar systems for instance, frequencies up to the W-band (75-110\,GHz) will be used. With these increased frequencies, new system architectures such as antenna-on-chip will be used, a much higher level of integration can be achieved \cite{7762824,5966314}, and the miniaturization of the complete system is enabled. Additionally, much higher absolute bandwidths can be achieved \cite{7762824}, resulting in increased data rates for telecommunication systems \cite{freemanTelecommunicationSystemEngineering2004} or improved range resolution for radar systems \cite{patoleAutomotiveRadarsReview2017}. 


Experimental demonstrations of transmissive metasurfaces operating in the W-band have been scarcely reported, despite the application potential \cite{pfeifferMillimeterWaveTransmitarraysWavefront2013,pfeifferBianisotropicMetasurfacesOptimal2014}. Refracting metasurfaces operating in the W-band were demonstrated by Pfeiffer et al. in 2013 \cite{pfeifferMillimeterWaveTransmitarraysWavefront2013}. More recent design techniques to reduce performance degradation and experimental methods which fully characterize all scattered energy have not yet been applied to this frequency range. There are several challenges, which make it more difficult to design efficient metasurfaces at mm-wave and to characterize them accurately. As metasurfaces are often designed with highly subwavelength elements, scaling down existing geometries such as those reported in Refs. \cite{epsteinHuygensMetasurfacesEquivalence2016,jiaLowreflectionBeamRefractions2015} results in feature sizes which are too small to be manufactured with standard PCB technology \cite{PhysRevApplied.11.064007}. Therefore, a significant effort has been undertaken to develop alternative concepts such as metasurfaces manufactured by screen-printing  \cite{wangSystematicDesignPrintable2018}. This method is limited to circuits with a single structured metallic layer, e.g. absorbing or reflecting metasurfaces. Comparably thin mm-wave lenses manufactured from perforated dielectrics have been proposed, however they require very complex fabrication processes \cite{imbertDesignPerformanceEvaluation2015,heMatchedLowLossWideband2018}. 


Periodic metasurfaces with very coarse discretization, also referred to as metagratings, have received attention and have been investigated mainly for anomalous reflection \cite{popovControllingDiffractionPatterns2018,rabinovichAnalyticalDesignPrintedCircuitBoard2018,wongPerfectAnomalousReflection2018,radiMetagratingsLimitsGraded2017,rabinovichExperimentalDemonstrationIndepth2019}. Some of these metagratings feature very simple geometric structures \cite{wongPerfectAnomalousReflection2018,radiMetagratingsLimitsGraded2017}. Therefore, such metagratings are easier to manufacture and they could potentially be scaled to higher frequencies like mm-wave or terahertz \cite{wongPerfectAnomalousReflection2018}. Most works on printed circuit metagratings \cite{popovControllingDiffractionPatterns2018,rabinovichAnalyticalDesignPrintedCircuitBoard2018}, including one on anomalous refraction \cite{rabinovichArbitraryDiffractionEngineering2019}, feature resonant elements based on thin capacitively loaded wires.  Here, the number of elements per period is reduced but the feature sizes of each resonant element are not significantly changed. Accordingly, demonstrations of transmissive printed circuit metagratings have also been limited to operating frequencies below 40\,GHz. 

From the perspective of ease of fabrication, a minimal number of large meta-atoms with relatively large feature size is preferred. However, we recently showed that the use of large resonators (cell size $\sim\lambda/3$) for transmissive metasurfaces can lead to significant near-field coupling effects within a cell \cite{PhysRevApplied.11.064007}. The perturbations caused by such near-field coupling effects were mitigated in previous works through numerical optimization \cite{ColeRefractionefficiencyHuygens2018}. These findings motivated the development of an improved synthesis method that incorporates such coupling effects and enables the efficient design of transmissive printed circuit metasurfaces for higher frequency applications \cite{PhysRevApplied.11.064007}. 
While the theoretical framework proposed in Ref. \cite{PhysRevApplied.11.064007} provides a basis for the synthesis of efficient mm-wave metasurfaces, several challenges in the circuit board design and the experimental characterization remain. At mm-wave frequencies, fabrication tolerances \cite{fischer_causes_2013}, the choice of circuit board materials \cite{zelenchuk_millimeter-wave_2012}, and conductor coatings \cite{usta_effects_2019} play an important role. Additionally, it was shown that the losses introduced by conductor roughness can significantly influence the performance of  mm-wave devices \cite{holmberg_surface_2018,curranMethodologyCombinedModeling2010,yong-hoon_kim_scattering_1999,goldPhysicalSurfaceRoughness2017}. New methodologies are available to include this effect into full-wave simulations \cite{goldPhysicalSurfaceRoughness2017}. 

Considering the amount of detail that needs to be modeled, good agreement between numerical simulation and experiment is important for the development of efficient mm-wave devices \cite{fischer_causes_2013}. We note that a metasurface which implements anomalous refraction is closely related to a blazed grating \cite{LaroucheReconciliationgeneralizedrefraction2012}, where all energy should be directed into a single diffraction order (or equivalently, Floquet harmonic).
In early experimental demonstrations, the energy scattered into each Floquet harmonic was only characterized at two or three different frequencies including the design frequency \cite{jiaLowreflectionBeamRefractions2015,wong_design_2014}.
Alternatively, the energy scattered into the direction of Floquet harmonics at the design frequency was measured versus frequency \cite{chen_theory_2018}, e.g. at 0$^\circ$ and $\pm$72$^\circ$ in this case. In recent works, the intensity of Floquet harmonics has been extracted using near-field scanning \cite{lavigne_susceptibility_2018,rabinovichArbitraryDiffractionEngineering2019} or using a bistatic far-field setup for reflective metagratings \cite{rabinovichExperimentalDemonstrationIndepth2019}. 


In this paper, highly efficient Huygens metasurfaces for anomalous beam refraction at 83\,GHz are designed and fabricated. We overcome previous performance limitations by using a synthesis procedure that follows  Ref. \cite{PhysRevApplied.11.064007} and compensates for inter-layer near-field coupling in an iterative optimization process. %
Additionally, losses caused by the surface roughness of conductors are included and their influence is minimized accordingly. The performance of our metasurfaces is demonstrated using an accurate determination of all forward scattering Floquet harmonics in a frequency range from 70-95\,GHz. We use a bistatic scattering cross section facility that is located in an anechoic chamber \cite{olkHighlyAccurateFullypolarimetric2017}, to determine the scattered far-field with a high angular precision and dynamic range. In order to ensure that the physics of the system is modeled accurately, we present a detailed analysis of our samples, including profilometry and microscopic imaging of polished cross-sections. We observe very good agreement between numerical simulation and experiment. Consequently, the presented experimental data verifies our numerical model.

\section{Design of mm-wave metasurfaces}

In this section, we show the architecture of the cells and macroscopic design considerations for Huygens metasurfaces exhibiting anomalous refraction.

\subsection{Cell architecture}

\begin{figure}[htbp]
	\centering
	\includegraphics[width=0.49\textwidth]{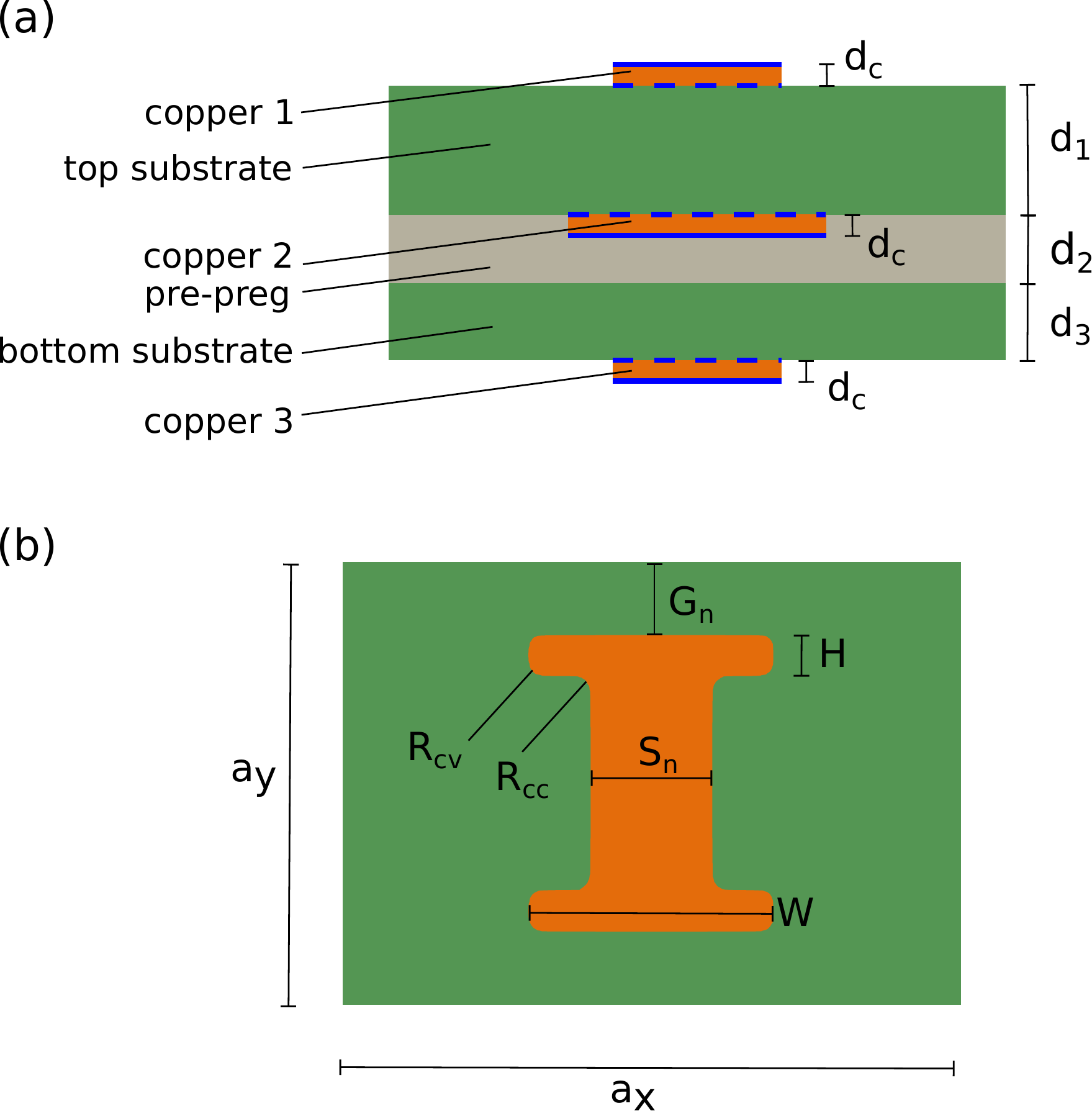}
	\caption{Circuit board layer stack-up (a) and layout of the metallic patterns (b).}
	\label{fig_layout}
\end{figure}

Every cell of the metasurfaces used in this work is synthesized according to the procedure presented in Ref. \cite{PhysRevApplied.11.064007} in order to fulfill a specific complex transmission $S_{21}$. In our case, all cells require near unity transmission amplitude. The transmission phase varies from cell to cell and depends on the refraction angle, as detailed in Subsection \ref{sec_supercell}. The design frequency is 80\,GHz. Our synthesis method incorporates an equivalent circuit model, which is fed with data from numerical simulations. In particular, the sheet impedance $Z_n$ of every metallic layer $n$ as well as corresponding near-field coupling coefficients are determined in full-wave simulations. 
The computational efficiency of our iterative synthesis algorithm is higher than black box optimization, and it is possible to consider complex shaped metallic patterns. Additionally, this procedure allows us to simulate the losses introduced by the conductor roughness by using the gradient surface model reported in Ref.~\cite{goldPhysicalSurfaceRoughness2017}. 
\markup{Utilizing this method, rough copper surfaces are modeled with effective conductivity and permeability in full-wave simulations.}

Losses can reduce the efficiency of microwave metasurfaces for two reasons. First, the efficiency is reduced simply because energy is absorbed in the structure. For a given set of material parameters and surface roughness, this part of performance degradation is unavoidable. The second reason is that with a poor estimate of the losses, the resonance frequency and quality factor are not accurately determined and the impedance is mismatched. 
In the particular case of refracting mm-wave metasurfaces, we found from full-wave simulation that the frequency shift caused by losses on rough conductors ranges from 1\,GHz to 2\,GHz \markup{(for further details see Subsection \ref{sec_supercell})}. Studies on other resonating microwave devices confirm the occurrence of such frequency shifts \cite{gopalakrishnanStudyEffectSurface2016}. This source of efficiency degradation is minimised here, as conductor losses and dielectric losses are taken into account in the numerical full wave simulation that feeds the optimization. For all numerical simulations in this work, we use the commercial software CST Microwave Studio \cite{cst}.  

The layer stack-up used here consists of three structured copper layers separated by dielectrics as shown in Figure \ref{fig_layout} (a). To achieve a nearly equal separation between all copper layers, we set the thickness of dielectric materials as $d_1 \approx d_2+d_3 \approx 254\,\upmu$m. The substrate and pre-preg are Isola Astra MT with a relative permittivity of $\epsilon_r=$3.00 and low dielectric loss tangent $\tan \delta=$0.0017. Given that the etching tolerance scales linearly with the copper thickness, we choose copper thickness $d_c=18\,\upmu$m, the smallest standard thickness available.
 
Special consideration was given to making the metallic pattern compatible with standard PCB manufacturing processes. As in previous works at lower frequencies \cite{lavigne_susceptibility_2018,capolino_equivalent_2013,rabinovichAnalyticalDesignPrintedCircuitBoard2018}, we use resonators in the shape of a dogbone. The geometric parameters are shown in Figure \ref{fig_layout}(b). We use a coarse discretization, with large cell size $a_x\approx\lambda/2.5$ and $a_y\approx\lambda/3.5$. All edges are rounded, which simplifies accurate manufacturing using photolithography. The radius of curvature is $R_{cv}=30\,\upmu$m for convex shapes and $R_{cc}=40\,\upmu$m for concave shapes. Additionally, all line and gap widths are chosen $\geq$ 100\,$\upmu$m, e.g. $H=100\,\upmu$m, $S_n=100...300\,\upmu$m and $G_n\geq90\,\upmu$m. 

The copper tracks of a circuit board which are in contact with air are usually coated with a submicron protection layer that prevents oxidation.  For mm-wave applications, the choice of the coating material is important as it can introduce significant losses. Here, a coating suitable for high frequency applications, immersion silver, was used. The conductivity of this coating is similar to copper and it has been shown to have only a small impact on conductor losses \cite{usta_effects_2019}. Therefore, we do not consider it in the numerical simulations.

\markup{The surface roughness on the two sides of each copper layer usually differ significantly, since they are determined by different steps in the manufacturing process \cite{goldPhysicalSurfaceRoughness2017}. The copper roughness on the side oriented towards the substrate $R_{q,s}$, marked with a dashed blue line in Figure \ref{fig_layout} (a), is usually higher than the roughness on the side towards the air or the pre-preg $R_{q,a}$, marked with a solid blue line. For the synthesis procedure, we use the PCB manufacturer's specifications of root mean square (RMS) roughness $R_{q,a}=0.4\,\upmu$m and $R_{q,s}=2.0\,\upmu$m, and DC conductivity 5.8${\times}10^7\,$S/m. The corresponding effective losses are determined with the methods from Ref.~\cite{goldPhysicalSurfaceRoughness2017}, and are shown in Figure \ref{fig_surfaceImp} in Appendix \ref{sec:lookup}.}

The geometrical parameters $S_n$ and $G_n$ are used to control the sheet impedance $Z_n$ of every metallic layer $n$. This dependence is characterized using numerical full-wave simulations and it is summarized in Figure \ref{fig_lookup} in Appendix \ref{sec:lookup}. For Huygens type metasurfaces, a symmetric stack-up of sheet impedances $Z_n$ is required. This means $Z_1=Z_3$ and we choose $G_1=G_3$ and $S_1=S_3$ accordingly. As there is no unique solution for the parameters $(G_n,S_n)$ to realize a specific $Z_n$, we choose those combinations that make the structure insensitive to parameter changes. More precisely, for each positive and negative value of $Z_n$ and for every $n$, we choose a constant value of $S_n$. This $S_n$ enables a large range of $\mathrm{imag}(Z_n)/\eta_0$ with variation of $G_n$ while maintaining a moderate slope $\frac{d Z_n}{d G_n}$, which minimizes sensitivity to fabrication errors.

\subsection{Supercell design \& refraction angles}
\label{sec_supercell}

Three periodic Huygens metasurfaces were designed and manufactured with different phase gradients in the $x$ direction. These correspond to 3, 4 and 5 different cells per period, i.e. the supercell widths $d_s^x$ are $3 a_x, 4 a_x$ and $5 a_x$. These designs are referred to with ``sample~1", ``sample~2" and ``sample~3", respectively. 

The required transmission response $S_{21}$ for each cell is chosen for near unity transmission amplitude and a phase that varies over $2\pi$ within one supercell, as per the procedure reported in Ref.~\cite{epsteinHuygensMetasurfacesEquivalence2016}. As an anomalously refracting Huygens metasurface cannot be perfectly matched to both the incident and transmitted waves, we choose to match  to the refracted wave. The corresponding propagating Floquet harmonics can be determined using the generalized law of refraction \cite{yu_light_2011}:
\begin{equation}
\label{eq_floquet_angle}
\sin(\theta_{out}^m) - \sin(\theta_{in}) = \frac{m \lambda}{d_s^x},
\end{equation}
where $\theta_{in}$ is the incident angle, $\theta_{out}^m$ is the refraction angle and $m$ the Floquet mode index. Modes are propagating when $|\sin(\theta_{out}^m)| < 1$. 

\begin{figure}[htbp]
	\centering
	\includegraphics[width=0.48\textwidth]{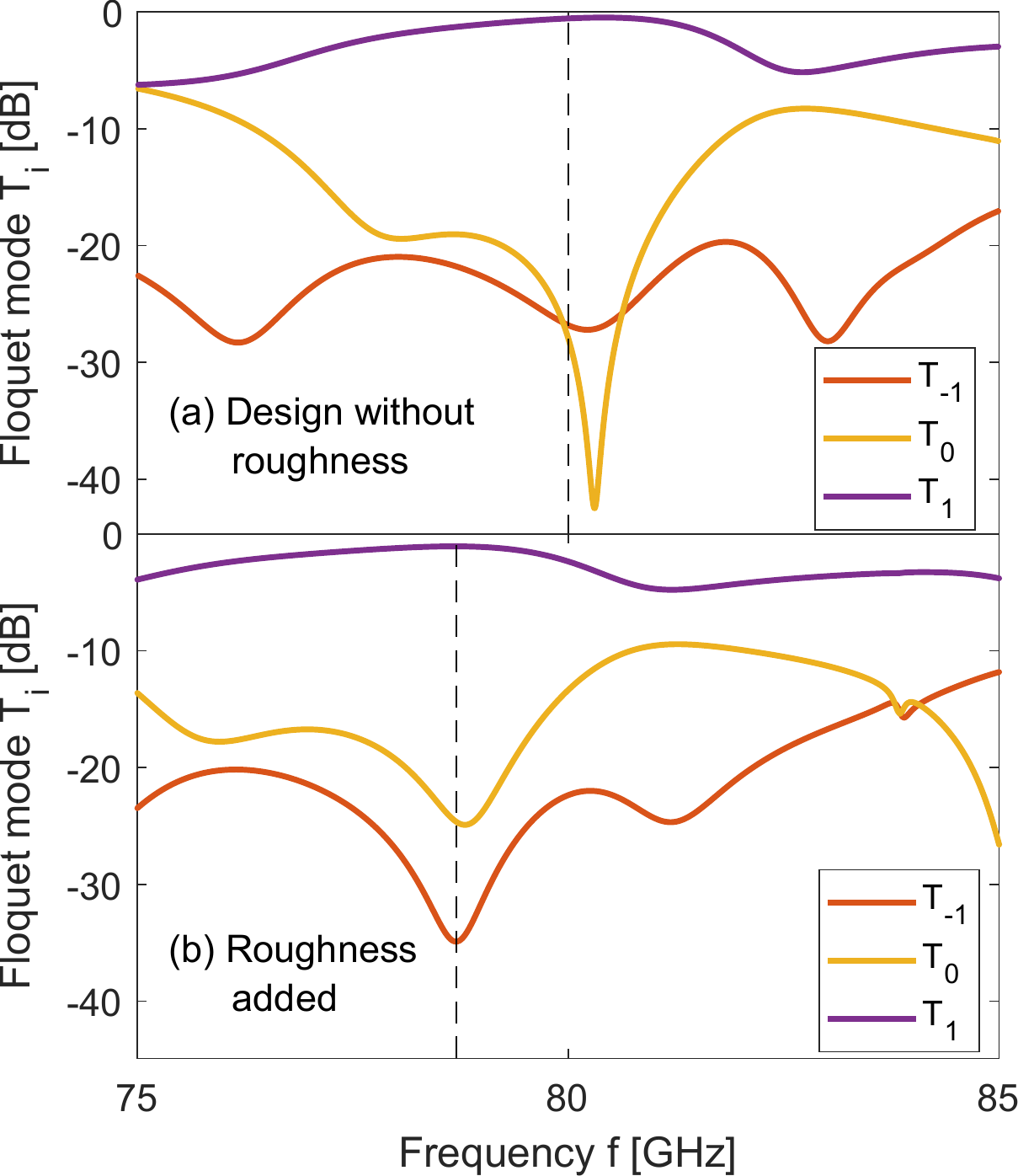}
	\caption{Numerically calculated performance of metasurface (a) neglecting copper roughness and (b) with copper roughness introduced after the synthesis procedure.}
	\label{fig_numExperiment}
\end{figure}

The samples were designed for normal incidence and the desired Floquet mode is $m=1$. The metasurface design aims to minimize the scattering into all other Floquet modes at the center frequency. In Table \ref{tab_geo} in Appendix \ref{sec:lookup}, we list the desired transmission phase $\angle(S_{21})$ as well as the initial normalized sheet reactances $\hat{X}_n=\mathrm{imag}(Z_n/\eta_0)$ for every cell of the design. Here, $\eta_0$ indicates the free space impedance. \markup{To demonstrate the significance of surface roughness, the synthesis procedure is first conducted for sample 2 ($\theta_{out}=38^\circ$ if $\theta_{in}=0$), neglecting surface roughness. The performance of this metasurface is shown in Figure \ref{fig_numExperiment} (a). Here, we observe a refraction efficiency of -0.5\,dB (88\,\%), with the remaining energy being partially scattered into undesired direction and partially absorbed. By analyzing power dissipation within each material, we find that 2.4\,\% of the incident energy is absorbed in dielectric materials, and 4.6\,\% is absorbed in the metal layers. In a second step, shown in Figure \ref{fig_numExperiment} (b), the same metasurface is analyzed with surface roughness turned on ($R_{q,a}=0.4\,\upmu$m and $R_{q,s}=2.0\,\upmu$m). The maximum refraction efficiency is reduced to -1.0\,dB (79\,\%), and it occurs 1.3\,GHz lower than the design frequency (dashed line). At this frequency, the undesired scattering and the dielectric losses are similar to the previous case, however absorption in the metal increases to 17\,\%. These results demonstrate that losses due to copper roughness dominate in our structure, and that failure to include these losses within the synthesis procedure will lead to a large shift in operating frequency. }

\markup{The iterative synthesis procedure is undertaken for each of our designed structures, including the effect of surface roughness. The resulting geometrical parameters $G_n$ and $S_n$ are shown in Table \ref{tab_geo} in Appendix \ref{sec:lookup} for every cell. We note that including the effect of copper roughness in the synthesis does not reduce the absorbed energy, but does eliminate the accompanying frequency shift. However, other fabrication imperfections, such as variances in the substrate thickness, under-etching or registration error between layers, can also significantly shift the frequency of optimal operation.  
The experimental characterization of sample 1, 2 and 3 is shown in Section \ref{sec_expChar}. Although these metasurface have been designed for 80\,GHz including the effects of copper roughness, the maximum refraction efficiency was found at 83\,GHz. The fabrication tolerances that cause this shift are measured and identified in Section \ref{sec_sample_analysis} and \ref{sec_comp}.}

\section{Experimental characterization}
\label{sec_expChar}

\subsection{Measurement of Floquet harmonics}

\begin{figure}[htbp]
	\centering
	\includegraphics[width=0.40\textwidth]{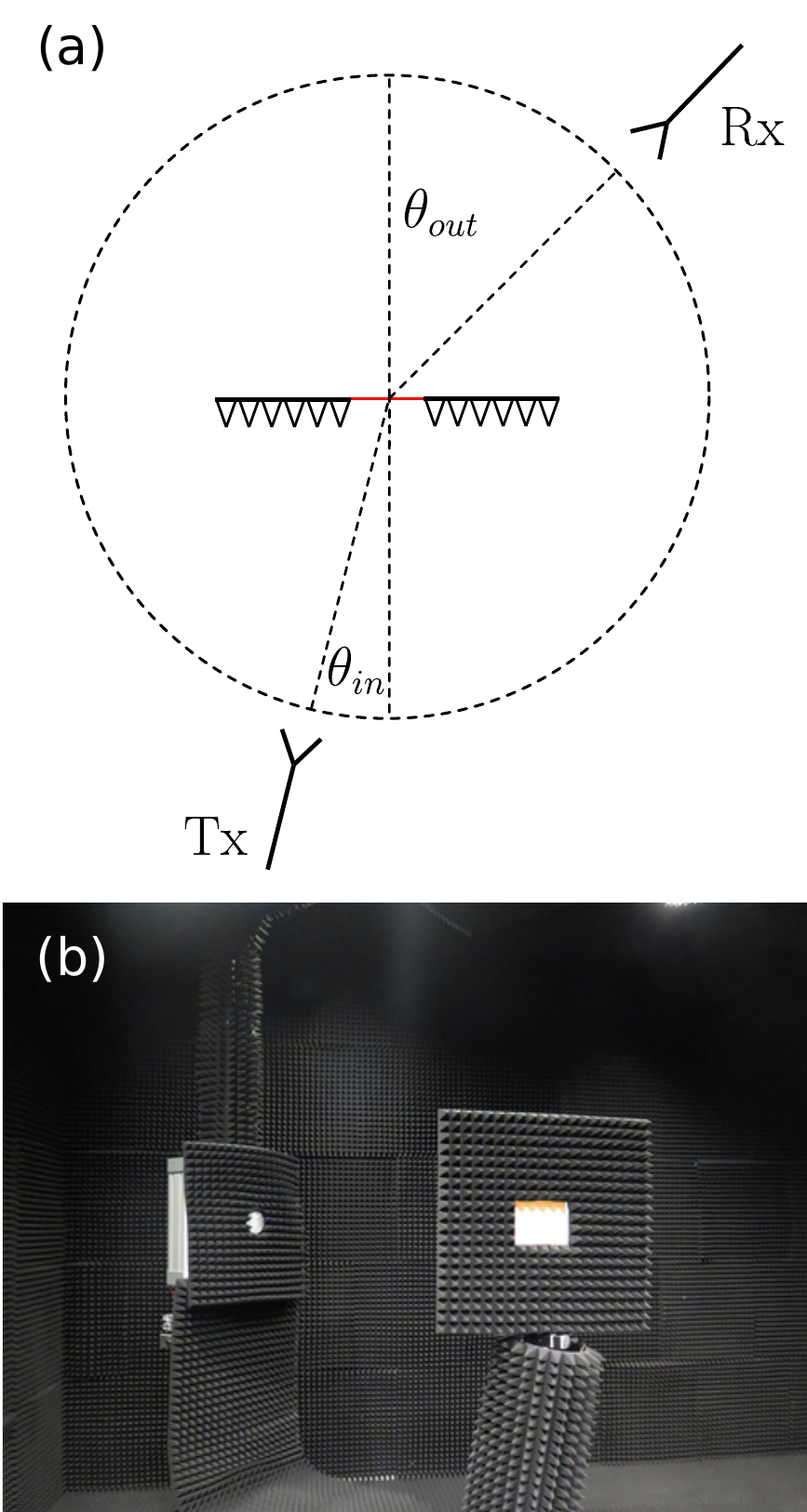}
	\caption{(a) Measurement configuration. (b) Photo of the setup including the sample support frame (right side) and the receiving antenna (left side).}
	\label{fig_exp_geo}
\end{figure}

The experimental characterization of propagating Floquet harmonics was conducted using a bistatic scattering cross section facility, which was reported in detail in Ref.~\cite{olkHighlyAccurateFullypolarimetric2017}. The measurement configuration is outlined in Figure \ref{fig_exp_geo} (a). Gaussian beam antennas were used, illuminating an area with full width half maximum diameter of approximately 50\,mm in the sample plane. The samples had a size of 140\,mm x 128\,mm and were mounted on a support frame of 400\,mm x 400\,mm. The samples were significantly larger than the illuminated area, therefore no edge effects are expected. Additionally, the support frame was covered with absorber as shown in Figure \ref{fig_exp_geo} (b). This minimizes spurious signals due to direct transmission between the test antennas and multiple reflections between the sample support and other objects in the measurement chamber. The bistatic system offers a positioning and alignment accuracy of $<0.1^\circ$. 

\begin{figure}[htbp]
	\centering
	\includegraphics[width=0.48\textwidth]{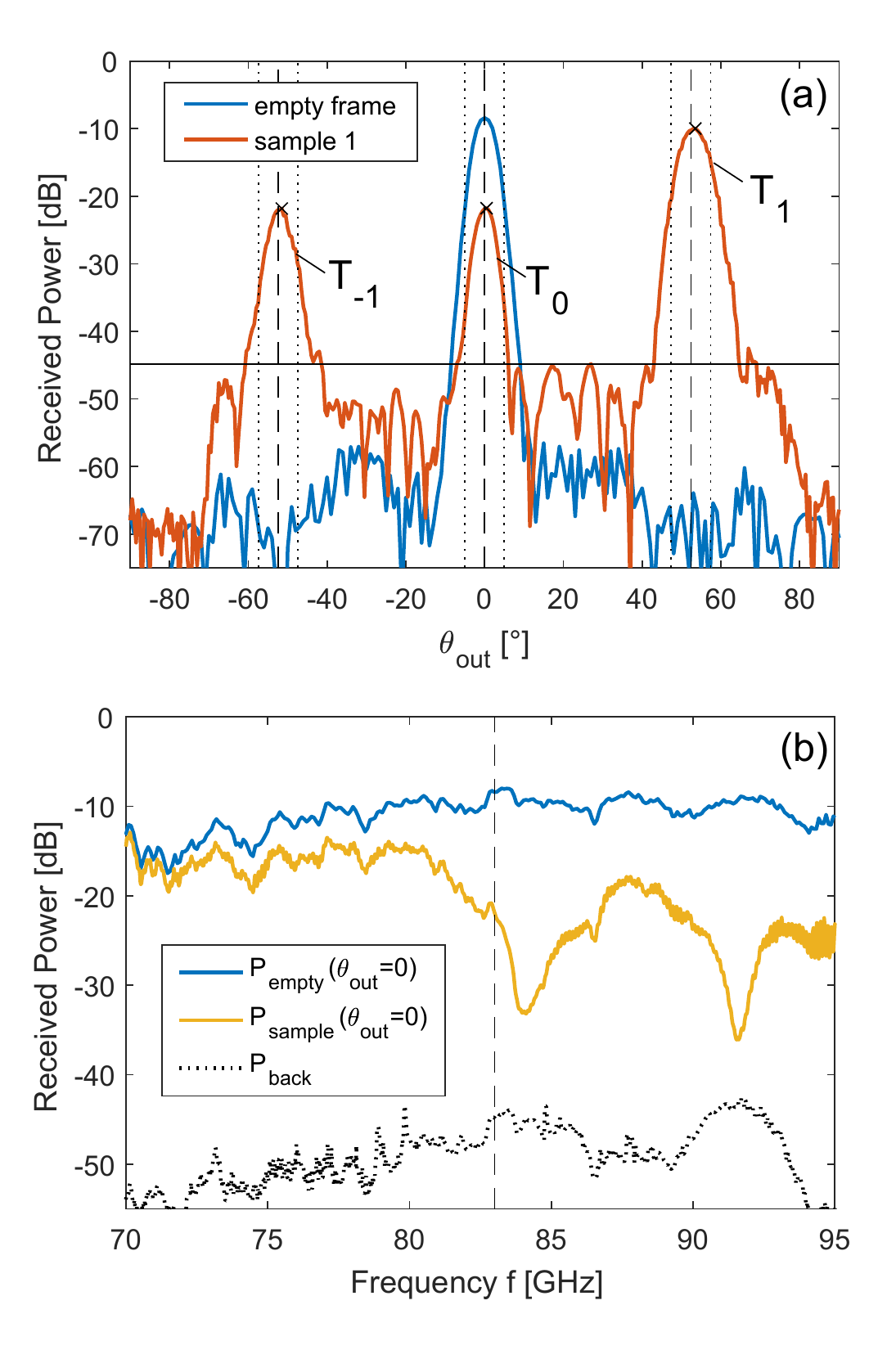}
	\caption{(a) Power scattered from sample 1 (red solid line) and from the empty sample support frame (blue solid line) at a frequency of 83\,GHz and an incident angle of $\theta_{in}=0$. (b) Normalization data from the empty frame measurement (blue) and peak power for the mode $T_0$ (yellow).}
	\label{fig_raw_data}
\end{figure}

The mm-wave system is based on a  Rohde \& Schwarz ZVA-67 vector network analyzer, with external frequency extenders and WR10 waveguide ports. To allow for movement of the receiving unit, long coaxial cables and several amplification stages are necessary. With a one-path through-open-short-match (TOSM) calibration \cite{hiebel_fundamentals_2011} using WR10 standards, the non-ideal response of these components was corrected. 
For each measurement we fixed the angle of incidence $\theta_{in}$ and swept the measurement angle $\theta_{out}$ from -90\,$^\circ$ to 90\,$^\circ$ in steps of 0.5$^\circ$. Measurements were taken over a frequency range from 70 to 95\,GHz \markup{and maximum efficiency was} found at 83\,GHz. The reasons for this discrepancy are discussed in Sections \ref{sec_sample_analysis} and \ref{sec_comp}. In Figure \ref{fig_raw_data} (a), the received power $P$ is plotted versus the scattering angle $\theta_{out}$ for a frequency of 83.0\,GHz at an incident angle of ${\theta_{in}=0^\circ}$. We compare the measurement of sample~1 (red) with the measurement of the empty sample support frame (blue). The measurement of the empty frame features one peak at $\theta_{out}=0^\circ$ whose width corresponds to the beam width of the test antenna set. The measurement of sample~1 features one main peak at $\theta_{out}=54.3^\circ$, which corresponds to the desired Floquet mode $T_1$. 

The intensity of the $T_1$ peak is only 1.3\,dB smaller than the empty frame main peak, indicating that 74\,\% of the incident energy is scattered into the desired direction. Due to the large dynamic range, the undesired Floquet modes $T_0$ and $T_{-1}$ are clearly visible. The peak position of each Floquet mode $m$ calculated from Eq.~\eqref{eq_floquet_angle} is indicated with a vertical black dashed line, and these lines agree with the maxima of the experimental measurements. We define a background scattering power level $P_{back}(f)$ as the maximum value of the $P_{sample}(f,\theta_{out})$ excluding the Floquet peaks. Here, $f$ indicates the frequency. This background scattering includes measurement artifacts such as side lobes and multiple scattering in the measurement chamber. In the case of sample~1 and $\theta_{in}=0^\circ$, $P_{back}(f=83\,$GHz$)$  is -45\,dB and is indicated by a solid horizontal line in Figure \ref{fig_raw_data} (a).

The frequency dependent intensity of Floquet harmonics is determined in two steps. First, each peak of the received power which corresponds to a Floquet harmonic $m$ needs to be evaluated throughout the frequency range of interest. One way to do this, would be to extract the received power $P_{sample}(f,\theta_{out}^{m})$ at the frequency dependent expected Floquet mode angle $\theta_{out}^{m}$ from Eq.~\eqref{eq_floquet_angle}. 
Due to experimental uncertainties, the measured peak position $\theta_{out}^{max,m}$ can slightly differ from this expected peak position $\theta_{out}^{m}$, leading to significant artifacts if the expected Floquet angle is taken.  Therefore, we choose to extract the received power at the measured peak position $\theta_{out}^{max,m}$. In particular, we determine $\theta_{out}^{max,m}(f)$ as the maximum in a region of $\pm5^\circ$ around the expected peak position $\theta_{out}^m$  (vertical black dotted lines) for each frequency sample ranging from 70 to 95\,GHz. The positions of these maximum values $\theta_{out}^{max,m}$ are indicated in Figure \ref{fig_raw_data} (a) with  black cross markers. 
In Figure \ref{fig_raw_data}~(b), the peak power $P_{sample}(f,\theta_{out}^{max,0})$ for Floquet mode $T_0$ is shown in comparison with the empty frame measurement $P_{empty}(\theta_{out}=0)$ and the background scattering $P_{back}$. The contrast between the empty frame measurement and the  scattering is on the order of 40\,dB and indicates the dynamic range in which Floquet modes can be measured. The received power $P_{sample}(f,\theta_{out}^{max,0})$ is significantly larger than the background scattering, confirming that the corresponding peak was detected reliably for all frequencies.

In the second step, the intensity of Floquet harmonics $|T_m|^2$ is obtained by normalizing the peak power values by the empty frame measurement, i.e. 
\begin{equation}
	|T_m(f)|^2 =  \frac{P_{sample}(f,\theta_{out}^{max,m})}{P_{empty}(f,\theta_{out}=0)}.
\end{equation}
With this normalization, the influence of the test antenna characteristics and the path losses are corrected for. We note that $T_m$ is a complex quantity. In this work, we consider only the absolute value $|T_m|^2$ and use it to evaluate the efficiency of periodic metasurfaces. An absolute value of $|T_m|^2=1$ for instance means that 100\% of the energy is scattered into Floquet mode $m$.

\subsection{Results}

\begin{figure}[htbp]
	\centering
	\includegraphics[width=0.48\textwidth]{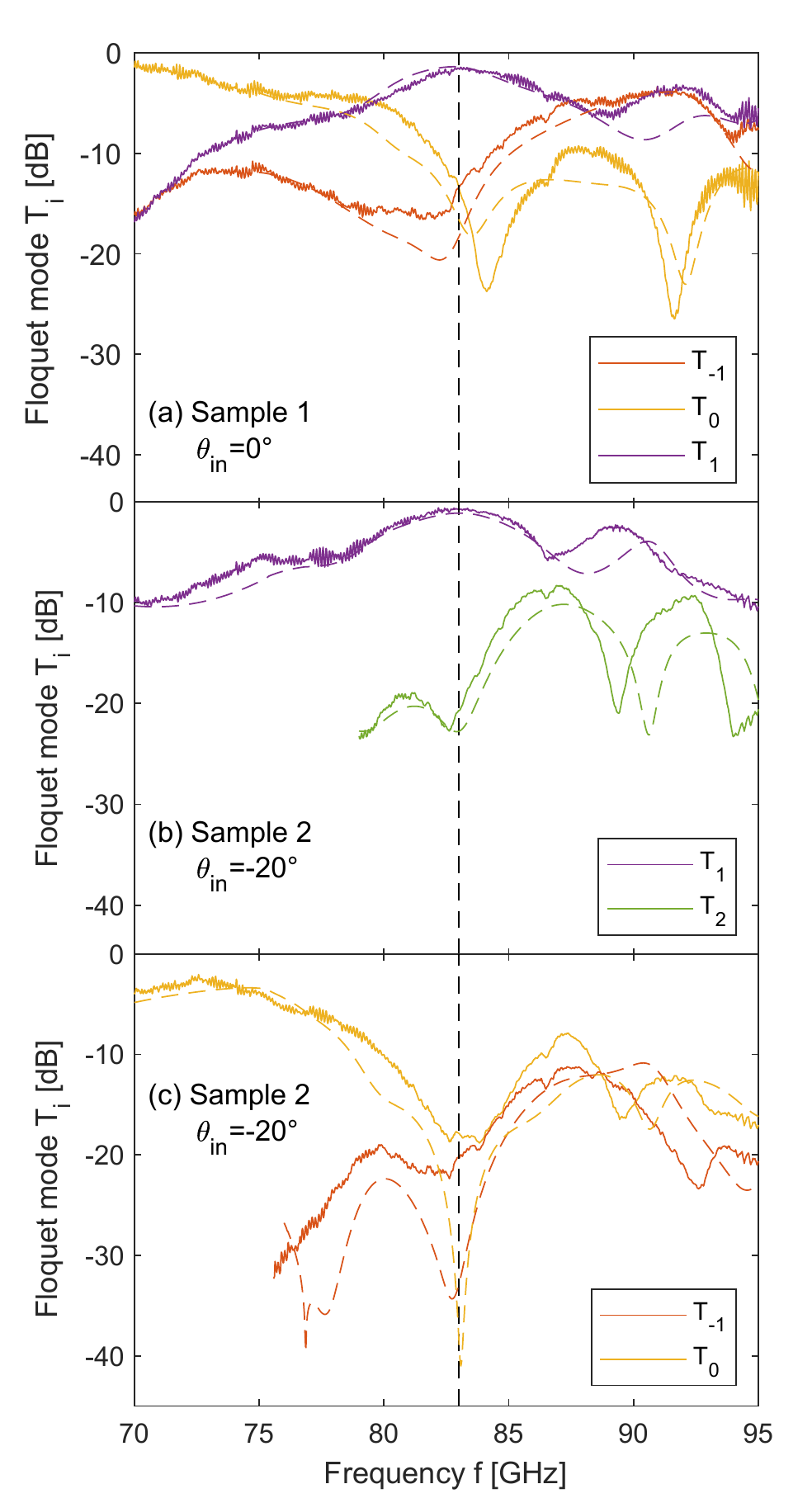}
	\caption{Floquet harmonics of refracting metasurfaces: Sample~1 (a) and sample~2 (b), (c). Solid line shows experimental values, dashed line shows results of full wave simulation.}
	\label{fig_result_plot}
\end{figure}

Each of the three samples is characterized at incidence angles $\theta_{in}$ of 0, -10$^\circ$ and -20$^\circ$. For this range of incident angles, each sample is relatively well impedance matched to the refracted wave. In each case, all Floquet harmonics propagating into the forward direction were measured over a frequency range from 70 to 95\,GHz. Throughout this frequency range, the background scattering $P_{back}(f)$ is always at least 7\,dB lower than the smallest Floquet mode peak $P_{sample}(f,\theta_{out}^{max,m})$ (typically 20\,dB lower).
The measurements of the first two samples at different incident angles are presented in Figure \ref{fig_result_plot}. The measurement of sample~1 excited at $\theta_{in}=0$ is shown in Figure \ref{fig_result_plot} (a) and sample~2 at $\theta_{in}=-20^\circ$ is shown in Figure \ref{fig_result_plot} (b) and (c). All other measurements are shown in Figure \ref{fig_result_large_plot} in Appendix \ref{sec:lookup}. Experimental curves are shown in solid lines. 

For sample~1 at $\theta_{in}=0^\circ$, we observe three propagating transmission modes, $T_{0}$ and $T_{\pm1}$. In case of sample~2 at $\theta_{in}=-20^\circ$, there are four propagating transmission modes, $T_{0}$, $T_{\pm1}$ and $T_{2}$. Here, $T_{-1}$ and $T_{2}$ are evanescent for frequencies below 75 and 77\,GHz, respectively. In the experiment, they can be observed for frequencies larger than 76 and 79~\,GHz,  which corresponds to angles $|\theta_{out}|$ smaller than about 80$^\circ$. In both cases, the $T_1$ mode is dominant at the design frequency, as expected. However, we observe a shift of the $T_1$ maximum from 80 to 83\,GHz. Other Floquet modes take comparably small values of -13.4 to -25\,dB at 83\,GHz.  

We summarize the beam refraction efficiency in Table \ref{tab_efficiency}. Here, we list the intensity of the desired Floquet mode $T_1$ at 83\,GHz which ranges from -1.6\,dB to -0.7\,dB (69 to 85\,\%). Additionally, we list the intensity of the strongest undesired Floquet mode $T_{ud}$, ranging from -13.4\,dB to -19.5\,dB. These findings will be compared with numerical simulations in section \ref{sec_comp}.

\begin{table}[]
	\centering
	\caption{Beam refraction efficiency}
	\label{tab_efficiency}
	\begin{tabular}{|c|c|ll|ll|}
		\hline
		\multirow{2}{.7cm}{sample} &  \multirow{2}{.35cm}{$\theta_{in}$}         & \multicolumn{2}{c|}{Experiment} & \multicolumn{2}{c|}{Simulation} \\
		&  & $T_1$ [dB] & $T_{ud}$ [dB] & $T_1$ [dB] & $T_{ud}$ [dB] \\ \hline
		1 & \multirow{3}{*}{0}                & -1.6       & -13.4        & -1.4       & -16.6        \\
		2 &                                   & -1.0       & -14.1        & -1.3       & -20.3        \\
		3 &                                   & -1.0       & -19.5        & -1.7       & -14.5        \\ \hline
		1 & \multirow{3}{*}{-10}              & -1.3       & -13.7        & -1.2       & -18.3         \\
		2 &                                   & -1.0       & -19.0        & -1.2       & -12.7        \\
		3 &                                   & -1.1       & -18.8        & -1.8       & -13.4        \\ \hline
		1 & \multirow{3}{*}{-20}              & -1.3       & -16.3        & -1.3       & -24.9        \\
		2 &                                   & -0.7       & -17.9        & -1.1       & -22.8        \\
		3 &                                   & -1.1       & -18.8        & -1.7       & -14.5       
		\\ \hline
	\end{tabular}
\end{table}

\subsection{Analysis of fabricated sample}
\label{sec_sample_analysis}

\begin{figure*}[h]
	\centering
	\includegraphics[width=0.98\textwidth]{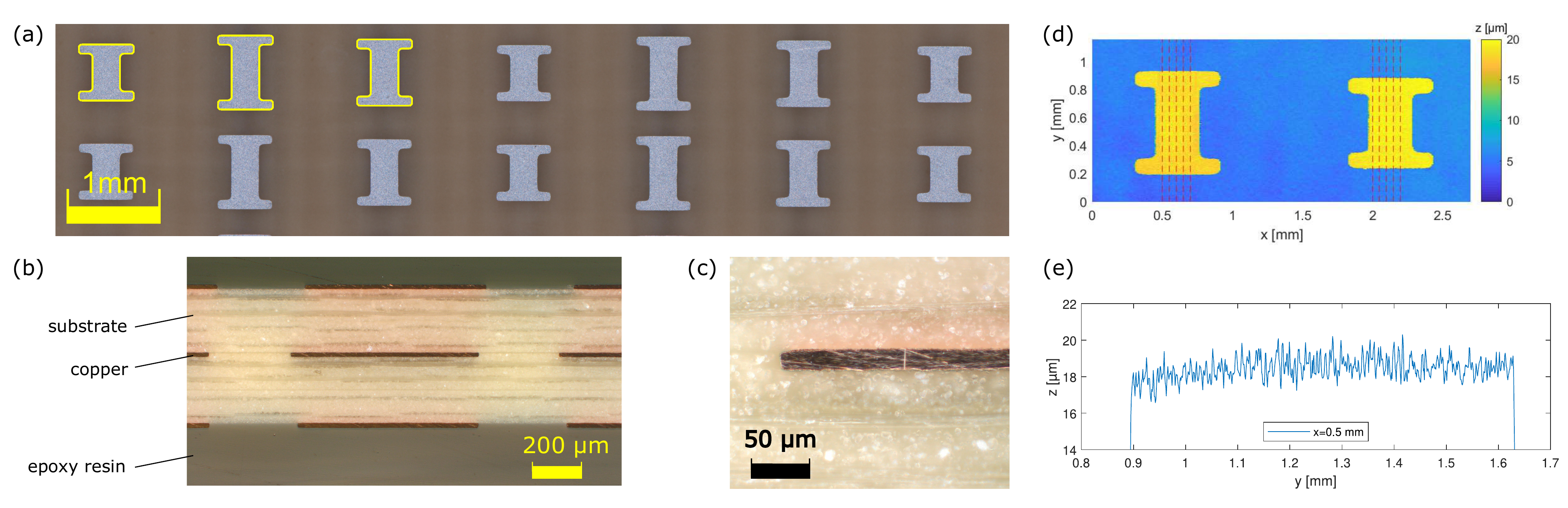}
	\caption{Analysis of circuit board samples: microscopy picture of the top surface (a), polished cross sections (b-c)  and optical profilometer scan (d-e).}
	\label{fig_sample}
\end{figure*}

In order to represent the geometry of our samples as accurate as possible in the numerical model, we perform a detailed analysis of the circuit boards. A high resolution picture of the top PCB surface shown in Figure \ref{fig_sample} (a) reveals the accuracy of the metallic patterns. We observe that the overall shape is etched very accurately, however it appears to be systematically smaller than the design. With an overlay of thin contour lines from the initial PCB layout (yellow), a scaling factor of $\alpha_s=0.97$ is determined.  We note that the scaling includes the shape of individual resonators as well as the distance from one to another, i.e. the periodicity of the array. Therefore, we assume that this inaccuracy stems from the photolithography or the mask preparation rather then from the etching process. 

The individual thicknesses of the layer stack-up are determined by analyzing polished PCB cross sections, as shown in Figure \ref{fig_sample} (b) and (c). The design thicknesses for the copper layers and for the top substrate are confirmed within the measurement accuracy of roughly 4\%, i.e. $d_c=$18$\upmu$m and $d_1=$254\,$\upmu$m. The pre-preg and the bottom substrate cannot be distinguished after fabrication, therefore we specify the sum $d_2+d_3=$ 265\,$\upmu$m.

\markup{A surface scan} of copper layer~1 that includes two metallic resonators is shown in Figure \ref{fig_sample} (d). This scan was performed using a step width of $\delta x = 10\upmu$m and $\delta y=1\upmu$m. We evaluate \markup{the RMS roughness} from several line scans in the y direction, which are marked with a dashed red line in Figure \ref{fig_sample} (d) and find $R_{q,a}=0.65\,\upmu$m. The line scan at $x=0.5\,$mm is shown separately in Figure \ref{fig_sample} (e). \markup{The roughness on the side of the substrate is determined approximately using cross sectional images such as Figure \ref{fig_sample} (b) and (c). This confirms the  manufacturers specification of $R_{q,s}=2\,\upmu$m.}

Additionally, the registration tolerance was analyzed using cross section images. According to the design, all resonators in one cell are concentric. We found however lateral shifts $\delta_l$ in $x$ and $y$ direction of up to 70\,$\upmu$m. Considering that these shifts are comparable to the geometrical dimensions, a significant perturbation was expected. Additionally, this layer misalignment varies from one supercell to the next, therefore it is impossible to include it into the numerical simulation with periodic boundaries. However, as shown below, good agreement can be achieved between numerical and experimental results, even when ignoring this effect.

\subsection{Comparison with simulation}
\label{sec_comp}

\begin{figure*}[htbp]
	\centering
	\includegraphics[width=0.99\textwidth]{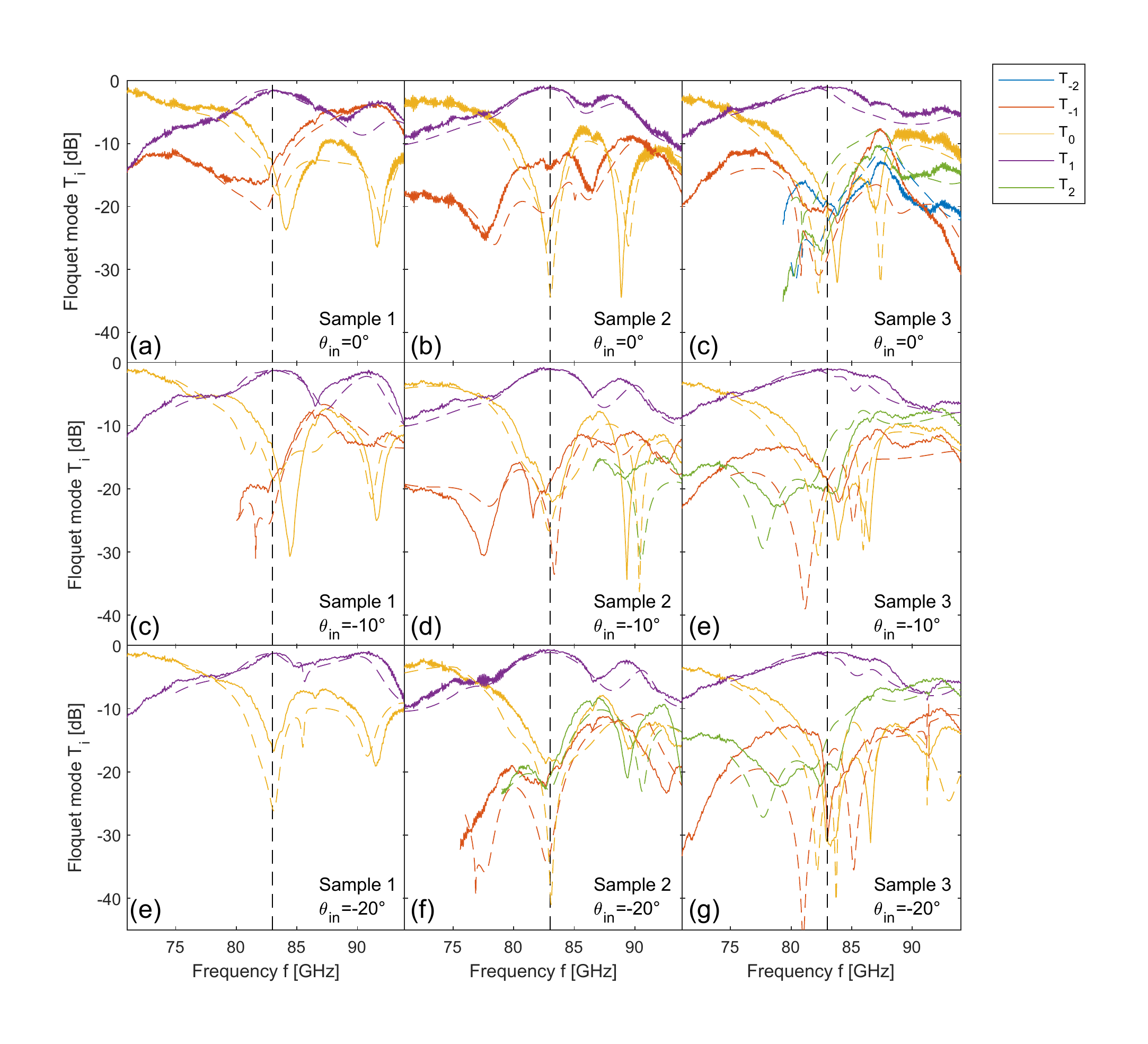}
	\caption{Floquet harmonics of all refracting metasurfaces fabricated in this work. Experiment (solid line) and full wave simulation (dashed lines) in comparison.}
	\label{fig_result_large_plot}
\end{figure*}

Besides of the registration tolerance, all geometrical parameters mentioned in subsection \ref{sec_sample_analysis} were used to refine the full wave simulation. As shown in Figures \ref{fig_result_plot} and \ref{fig_result_large_plot}, there is a good agreement between experiment (solid lines) and simulation (dashed lines), especially for the desired Floquet mode $T_1$. For the undesired modes, the main feature of the experimental and numerical curves match. Large discrepancies only occur for values below -20\,dB.

As the geometrical parameters were updated individually to match the sample measurement, it was observed that the systematic scaling of the metallic pattern described by the scaling factor $\alpha_s$ is the main cause of the observed frequency shift. If this scaling error is consistent from one manufacturing run to the next, in principle it could be compensated for. Variations in the copper roughness and the substrate thickness are rather impacting secondary features in the Floquet modes such as the depth and the position of local minima. 

\begin{figure}[htbp]
	\centering
	\includegraphics[width=0.49\textwidth]{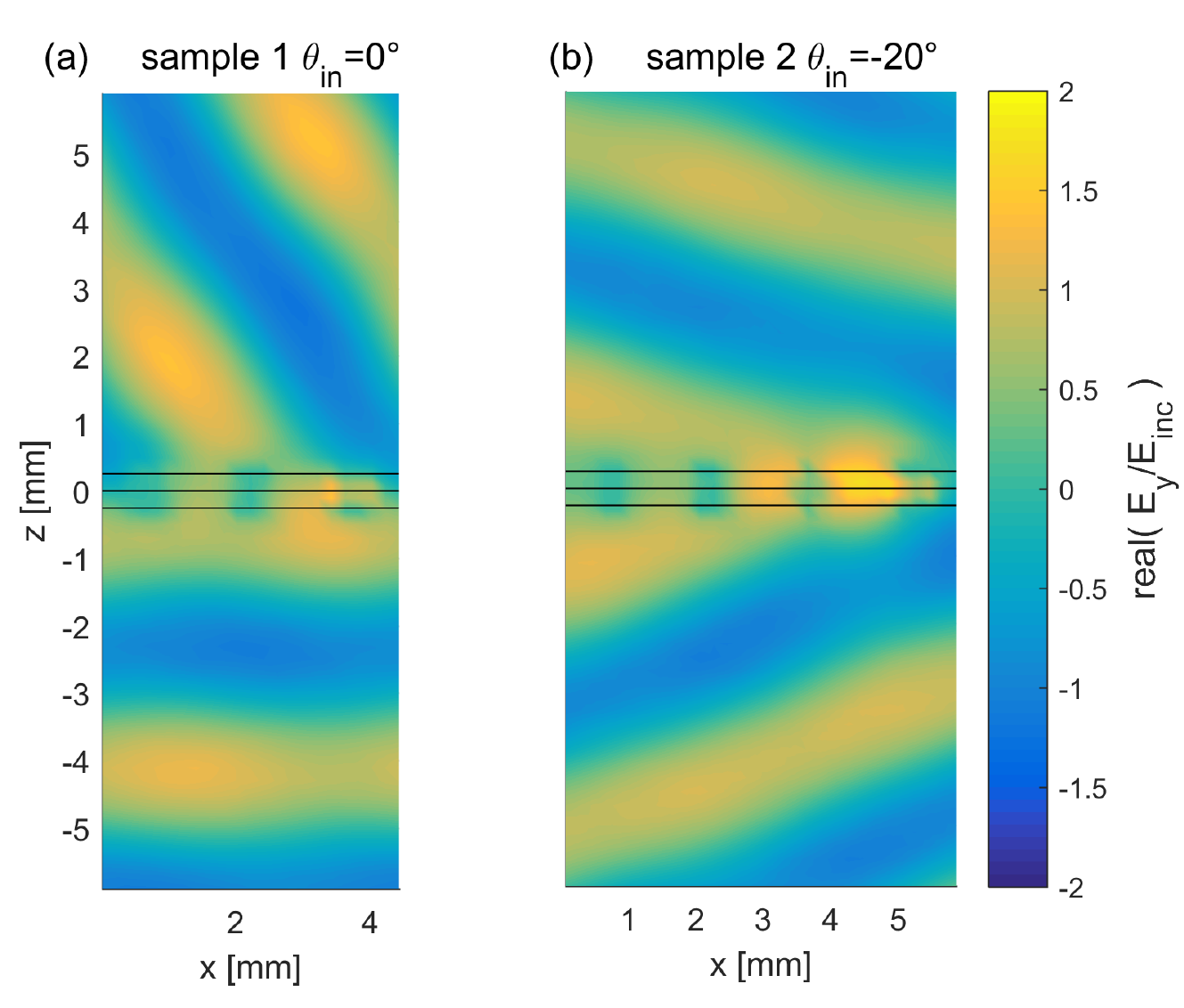}
	\caption{Numerically simulated electric field in the plane $y=0$ propagating through (a) sample 1 at $\theta_{in}=0^\circ$ and (b) sample 2 at $\theta_{in}=-20^\circ$.}
	\label{fig_field}
\end{figure}

Given that the reliability of the numerical simulation has been verified, further analysis that is inaccessible in the experiment can be performed numerically. A plot of the electric field inside and close to the metasurfaces for instance is not only very illustrative, it can help to identify defects in the design of individual cells of the metasurface. In Figure \ref{fig_field}, we show numerically calculated electric field distributions for sample\,1 and sample\,2. In both cases, the electric field features relatively planar wavefronts propagating into the desired direction, confirming that good transmission was achieved through all cells.


\section{Conclusion}

Highly efficient refracting metasurfaces for the mm-wave frequency range are presented. The samples are designed for fabrication with standard PCB processes, which is a significant challenge for this frequency range. We use an advanced synthesis procedure that accounts for the most significant causes of performance degradation, including near-field coupling and the losses caused by conductor roughness. The experimental characterization and data processing technique used here  allow for a very accurate determination of the intensity of Floquet harmonics over a wide frequency range. A shift of the optimal operation frequency from 80 to 83\,GHz is observed. According to a profilometer and microscopy analysis of the samples, this shift can be attributed to the tolerance of the photolithography.

Beam refraction efficiencies on the order of -1.0\,dB (80\,\%) were achieved, showing good agreement between experiment and numerical simulation. This outcome verifies our numerical model and shows the relevance of incorporating secondary effects such as surface roughness into the metasurface synthesis. Additionally, given that the model was verified quantitatively, information inaccessible in the experiment can be calculated numerically with high reliability.  Furthermore, the efficiency prediction for future studies which are done with the same fabrication process can potentially be improved.


%


\section*{Acknowledgment}

This work was financially supported by the Australian Research Council (Linkage Project LP160100253), the  Luxembourg Ministry of the Economy (grant CVN 18/17/RED) and the University of New South Wales (UIPA scholarship).

\newpage
\appendices
\section{} \label{sec:lookup}

\begin{figure}[h]
	\centering
	\includegraphics[width=0.42\textwidth]{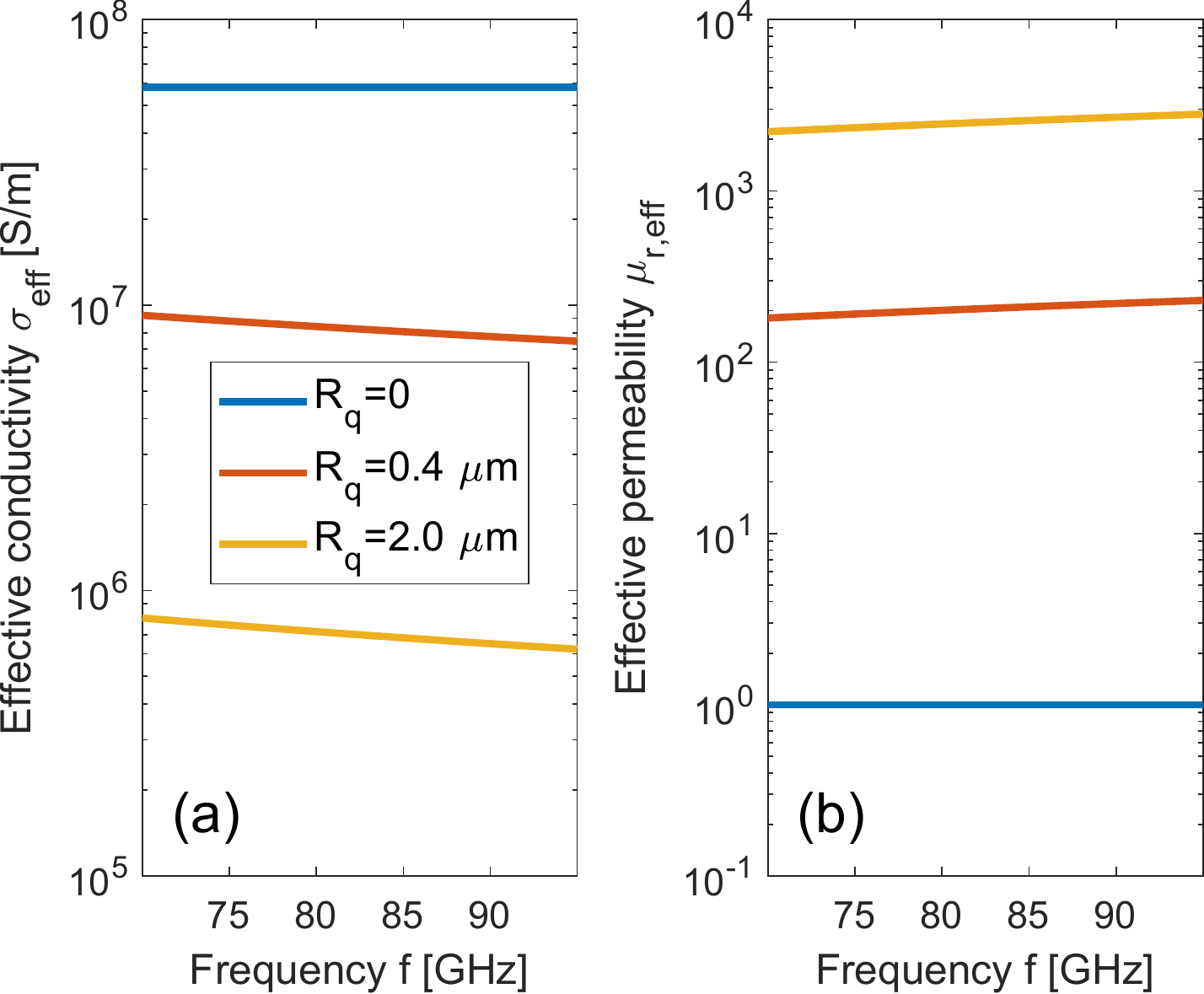}
	\caption{Effective conductivity $\sigma_{eff}$ (a) and effective permeability $\mu_{r,eff}$ (b) of a copper surface with an RMS roughness $R_q$ of 0, 0.4\,$\upmu$m and 2.0\,$\upmu$m.}
	\label{fig_surfaceImp}
\end{figure}

\FloatBarrier

\begin{figure}[h]
	\centering
	\includegraphics[width=0.38\textwidth]{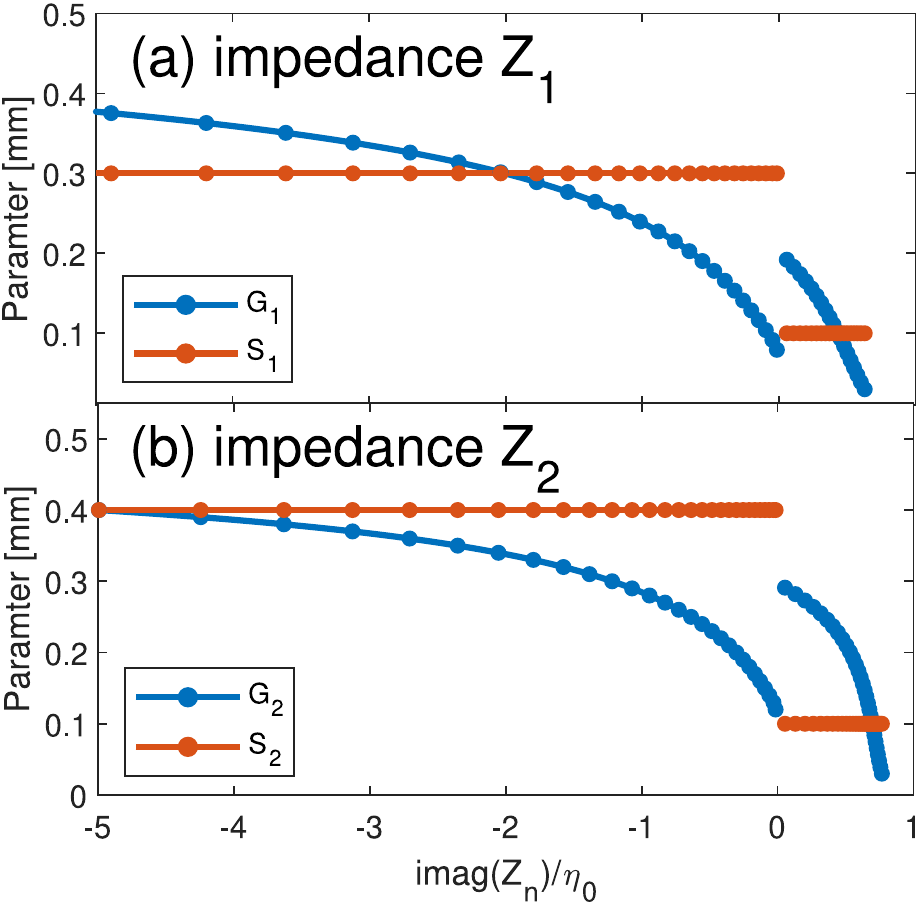}
	\caption{Lookup table relating the geometrical parameters $G_n$ and $S_n$ to the sheet impedance $Z_n$.}
	\label{fig_lookup}
\end{figure}

\begin{table}[h]
	\centering
	\caption{Design parameters}
	\label{tab_geo}
	\scalebox{0.9}{
	\begin{tabular}{|c|c|r|rr|cccc|}
		\hline
		\multirow{2}{*}{sample} & \multirow{2}{*}{cell}    & \multirow{2}{*}{$\angle S_{21}$ [$^\circ$]} & \multirow{2}{*}{$\hat{X}_1$} & \multirow{2}{*}{$\hat{X}_2$} &  \multicolumn{4}{c|}{Geometry [$\upmu$m] } \\
		&&&& & $G_1$                    & $S_1$                    & $G_2$                    & $S_2$                    \\ \hline
		\multirow{3}{*}{1} & 1&151&-0.49&-0.11    & 179                     & 300                     & 165                     & 400                     \\
		& 2&39&-0.27&-0.10    & 128                     & 300                     & 153                     & 400                     \\
		& 3&-86&-0.78&0.13    & 232                     & 300                     & 293                     & 100                     \\ \hline
		\multirow{4}{*}{2} & 1&163&-0.49&-0.16    & 182                     & 300                     & 174                     & 400                     \\
		& 2&73&-0.34&-0.08    & 147                     & 300                     & 150                     & 400                     \\
		& 3&-17&0.30&-0.74    & 147                     & 100                     & 276                     & 400                     \\
		& 4&-107&-0.76&0.24    & 228                     & 300                     & 278                     & 100                     \\ \hline
		\multirow{5}{*}{3} & 1&120&-0.42&-0.10    & 169                     & 300                     & 151                     & 400                     \\
		& 2&48&-0.26&-0.11    & 136                     & 300                     & 150                     & 400                     \\
		& 3&-24&0.46&-1.41    & 96                      & 100                     & 325                     & 400                     \\
		& 4&-96&-0.89&0.27    & 243                     & 300                     & 277                     & 100                     \\
		& 5&-168&-0.55&-0.45    & 191                     & 300                     & 224                     & 400                    \\ \hline
	\end{tabular}}
\end{table}

\ifCLASSOPTIONcaptionsoff
  \newpage
\fi



%
%
%


\bibliographystyle{IEEEtran}
\bibliography{mm-wave_beam_refraction}

\begin{thebibliography}{10}
\providecommand{\url}[1]{#1}
\csname url@samestyle\endcsname
\providecommand{\newblock}{\relax}
\providecommand{\bibinfo}[2]{#2}
\providecommand{\BIBentrySTDinterwordspacing}{\spaceskip=0pt\relax}
\providecommand{\BIBentryALTinterwordstretchfactor}{4}
\providecommand{\BIBentryALTinterwordspacing}{\spaceskip=\fontdimen2\font plus
\BIBentryALTinterwordstretchfactor\fontdimen3\font minus
  \fontdimen4\font\relax}
\providecommand{\BIBforeignlanguage}[2]{{%
\expandafter\ifx\csname l@#1\endcsname\relax
\typeout{** WARNING: IEEEtran.bst: No hyphenation pattern has been}%
\typeout{** loaded for the language `#1'. Using the pattern for}%
\typeout{** the default language instead.}%
\else
\language=\csname l@#1\endcsname
\fi
#2}}
\providecommand{\BIBdecl}{\relax}
\BIBdecl

\bibitem{yuFlatOpticsDesigner2014}
N.~Yu and F.~Capasso, ``\BIBforeignlanguage{en}{Flat optics with designer
  metasurfaces},'' \emph{\BIBforeignlanguage{en}{Nature Materials}}, vol.~13,
  no.~2, pp. 139--150, Feb. 2014.

\bibitem{glybovskiMetasurfacesMicrowavesVisible2016}
S.~B. Glybovski, S.~A. Tretyakov, P.~A. Belov, Y.~S. Kivshar, and C.~R.
  Simovski, ``Metasurfaces: {{From}} microwaves to visible,'' \emph{Physics
  Reports}, vol. 634, pp. 1--72, May 2016.

\bibitem{dingGradientMetasurfacesReview2018}
F.~Ding, A.~Pors, and S.~I. Bozhevolnyi, ``\BIBforeignlanguage{en}{Gradient
  metasurfaces: A review of fundamentals and applications},''
  \emph{\BIBforeignlanguage{en}{Reports on Progress in Physics}}, vol.~81,
  no.~2, p. 026401, Feb. 2018.

\bibitem{KAOUACH20161275}
H.~Kaouach, A.~Kabashi, and F.~Ouasli, ``High-efficiency wideband
  transmit-array antenna with linear polarization in {{Q}}-band,'' \emph{AEU -
  International Journal of Electronics and Communications}, vol.~70, no.~9, pp.
  1275--1281, 2016.

\bibitem{ryanWidebandTransmitarrayUsing2010}
C.~G.~M. Ryan, M.~R. Chaharmir, J.~Shaker, J.~R. Bray, Y.~M.~M. Antar, and
  A.~Ittipiboon, ``\BIBforeignlanguage{en}{A {{Wideband Transmitarray Using
  Dual}}-{{Resonant Double Square Rings}}},''
  \emph{\BIBforeignlanguage{en}{IEEE Transactions on Antennas and
  Propagation}}, vol.~58, no.~5, pp. 1486--1493, May 2010.

\bibitem{saeidiFigureMeritFocusing2015}
C.~Saeidi and D.~{van der Weide}, ``\BIBforeignlanguage{en}{A figure of merit
  for focusing metasurfaces},'' \emph{\BIBforeignlanguage{en}{Applied Physics
  Letters}}, vol. 106, no.~11, p. 113110, Mar. 2015.

\bibitem{chenDesignExperimentalVerification2019}
M.~Chen, A.~Epstein, and G.~V. Eleftheriades, ``Design and {{Experimental
  Verification}} of a {{Passive Huygens}}' {{Metasurface Lens}} for {{Gain
  Enhancement}} of {{Frequency}}-{{Scanning Slotted}}-{{Waveguide Antennas}},''
  \emph{IEEE Transactions on Antennas and Propagation}, pp. 1--1, 2019.

\bibitem{abdo-sanchezLeakyWaveAntennaControlled2018}
E.~{Abdo-Sanchez}, M.~Chen, A.~Epstein, and G.~V. Eleftheriades,
  ``\BIBforeignlanguage{en}{A {{Leaky}}-{{Wave Antenna With Controlled
  Radiation Using}} a {{Bianisotropic Huygens}}' {{Metasurface}}},''
  \emph{\BIBforeignlanguage{en}{IEEE Transactions on Antennas and
  Propagation}}, pp. 1--1, 2018.

\bibitem{liHolographicLeakywaveMetasurfaces2015}
Y.~B. Li, L.~L. Li, B.~G. Cai, Q.~Cheng, and T.~J. Cui,
  ``\BIBforeignlanguage{en}{Holographic leaky-wave metasurfaces for dual-sensor
  imaging},'' \emph{\BIBforeignlanguage{en}{Scientific Reports}}, vol.~5, p.
  18170, Dec. 2015.

\bibitem{pfeifferMetamaterialHuygensSurfaces2013a}
C.~Pfeiffer and A.~Grbic, ``Metamaterial {{Huygens}}' {{Surfaces}}: {{Tailoring
  Wave Fronts}} with {{Reflectionless Sheets}},'' \emph{Physical Review
  Letters}, vol. 110, no.~19, p. 197401, May 2013.

\bibitem{epsteinHuygensMetasurfacesEquivalence2016}
A.~Epstein and G.~V. Eleftheriades, ``Huygens' metasurfaces via the equivalence
  principle: Design and applications,'' \emph{Journal of the Optical Society of
  America B}, vol.~33, no.~2, pp. A31--A50, 2016.

\bibitem{jiaLowreflectionBeamRefractions2015}
S.~L. Jia, X.~Wan, X.~J. Fu, Y.~J. Zhao, and T.~J. Cui,
  ``\BIBforeignlanguage{en}{Low-reflection beam refractions by ultrathin
  {{Huygens}} metasurface},'' \emph{\BIBforeignlanguage{en}{AIP Advances}},
  vol.~5, no.~6, p. 067102, Jun. 2015.

\bibitem{wong_design_2014}
\BIBentryALTinterwordspacing
J.~P.~S. Wong, M.~Selvanayagam, and G.~V. Eleftheriades, ``Design of unit cells
  and demonstration of methods for synthesizing huygens metasurfaces,''
  \emph{Photonics and Nanostructures - Fundamentals and Applications}, vol.~12,
  no.~4, pp. 360--375, 2014. [Online]. Available:
  \url{http://www.sciencedirect.com/science/article/pii/S1569441014000686}
\BIBentrySTDinterwordspacing

\bibitem{7762824}
B.~Goettel, P.~Pahl, C.~Kutschker, S.~Malz, U.~R. Pfeiffer, and T.~Zwick,
  ``Active {{Multiple Feed On}}-{{Chip Antennas With Efficient In}}-{{Antenna
  Power Combining Operating}} at 200\textendash{}320 {{GHz}},'' \emph{IEEE
  Transactions on Antennas and Propagation}, vol.~65, no.~2, pp. 416--423, Feb.
  2017.

\bibitem{5966314}
S.~Pan and F.~Capolino, ``Design of a {{CMOS On}}-{{Chip Slot Antenna With
  Extremely Flat Cavity}} at 140 {{GHz}},'' \emph{IEEE Antennas and Wireless
  Propagation Letters}, vol.~10, pp. 827--830, 2011.

\bibitem{6415978}
T.~Shen, T.~J. Kao, T.~Huang, J.~Tu, J.~Lin, and R.~Wu, ``Antenna {{Design}} of
  60-{{GHz Micro}}-{{Radar System}}-{{In}}-{{Package}} for {{Noncontact Vital
  Sign Detection}},'' \emph{IEEE Antennas and Wireless Propagation Letters},
  vol.~11, pp. 1702--1705, 2012.

\bibitem{freemanTelecommunicationSystemEngineering2004}
R.~L. Freeman, \emph{Telecommunication {{System Engineering}}}, 3rd~ed.\hskip
  1em plus 0.5em minus 0.4em\relax {Wiley}, 2004.

\bibitem{patoleAutomotiveRadarsReview2017}
S.~M. Patole, M.~Torlak, D.~Wang, and M.~Ali, ``Automotive radars: {{A}} review
  of signal processing techniques,'' \emph{IEEE Signal Processing Magazine},
  vol.~34, no.~2, pp. 22--35, Mar. 2017.

\bibitem{pfeifferMillimeterWaveTransmitarraysWavefront2013}
C.~Pfeiffer and A.~Grbic, ``Millimeter-{{Wave Transmitarrays}} for
  {{Wavefront}} and {{Polarization Control}},'' \emph{IEEE Transactions on
  Microwave Theory and Techniques}, vol.~61, no.~12, pp. 4407--4417, Dec. 2013.

\bibitem{pfeifferBianisotropicMetasurfacesOptimal2014}
------, ``\BIBforeignlanguage{en}{Bianisotropic {{Metasurfaces}} for {{Optimal
  Polarization Control}}: {{Analysis}} and {{Synthesis}}},''
  \emph{\BIBforeignlanguage{en}{Physical Review Applied}}, vol.~2, no.~4, Oct.
  2014.

\bibitem{PhysRevApplied.11.064007}
A.~Olk and D.~Powell, ``Accurate {{Metasurface Synthesis Incorporating
  Near}}-{{Field Coupling Effects}},'' \emph{Phys. Rev. Applied}, vol.~11,
  no.~6, p. 064007, Jun. 2019.

\bibitem{wangSystematicDesignPrintable2018}
X.~Wang, A.~{D\'iaz-Rubio}, A.~Sneck, A.~Alastalo, T.~M\"akel\"a,
  J.~{Ala-Laurinaho}, J.~Zheng, A.~V. R\"ais\"anen, and S.~A. Tretyakov,
  ``Systematic {{Design}} of {{Printable Metasurfaces}}: {{Validation Through
  Reverse}}-{{Offset Printed Millimeter}}-{{Wave Absorbers}},'' \emph{IEEE
  Transactions on Antennas and Propagation}, vol.~66, no.~3, pp. 1340--1351,
  Mar. 2018.

\bibitem{imbertDesignPerformanceEvaluation2015}
M.~Imbert, A.~Papi\'o, F.~D. Flaviis, L.~Jofre, and J.~Romeu, ``Design and
  {{Performance Evaluation}} of a {{Dielectric Flat Lens Antenna}} for
  {{Millimeter}}-{{Wave Applications}},'' \emph{IEEE Antennas and Wireless
  Propagation Letters}, vol.~14, pp. 342--345, 2015.

\bibitem{heMatchedLowLossWideband2018}
Y.~He and G.~V. Eleftheriades, ``Matched, {{Low}}-{{Loss}}, and {{Wideband
  Graded}}-{{Index Flat Lenses}} for {{Millimeter}}-{{Wave Applications}},''
  \emph{IEEE Transactions on Antennas and Propagation}, vol.~66, no.~3, pp.
  1114--1123, Mar. 2018.

\bibitem{popovControllingDiffractionPatterns2018}
V.~Popov, F.~Boust, and S.~N. Burokur, ``\BIBforeignlanguage{en}{Controlling
  {{Diffraction Patterns}} with {{Metagratings}}},''
  \emph{\BIBforeignlanguage{en}{Physical Review Applied}}, vol.~10, no.~1, Jul.
  2018.

\bibitem{rabinovichAnalyticalDesignPrintedCircuitBoard2018}
O.~Rabinovich and A.~Epstein, ``\BIBforeignlanguage{en}{Analytical design of
  printed-circuit-board ({{PCB}}) metagratings for perfect anomalous
  reflection},'' \emph{\BIBforeignlanguage{en}{IEEE Transactions on Antennas
  and Propagation}}, pp. 4086--4095, 2018.

\bibitem{wongPerfectAnomalousReflection2018}
A.~M.~H. Wong and G.~V. Eleftheriades, ``Perfect {{Anomalous Reflection}} with
  a {{Bipartite Huygens}}' {{Metasurface}},'' \emph{Physical Review X}, vol.~8,
  no.~1, p. 011036, Feb. 2018.

\bibitem{radiMetagratingsLimitsGraded2017}
Y.~Ra'di, D.~L. Sounas, and A.~Al\`u, ``\BIBforeignlanguage{en}{Metagratings:
  {{Beyond}} the {{Limits}} of {{Graded Metasurfaces}} for {{Wave Front
  Control}}},'' \emph{\BIBforeignlanguage{en}{Physical Review Letters}}, vol.
  119, no.~6, Aug. 2017.

\bibitem{rabinovichExperimentalDemonstrationIndepth2019}
O.~Rabinovich, I.~Kaplon, J.~Reis, and A.~Epstein, ``Experimental demonstration
  and in-depth investigation of analytically designed anomalous reflection
  metagratings,'' \emph{Physical Review B}, vol.~99, no.~12, p. 125101, Mar.
  2019.

\bibitem{rabinovichArbitraryDiffractionEngineering2019}
O.~Rabinovich and A.~Epstein, ``\BIBforeignlanguage{en}{Arbitrary {{Diffraction
  Engineering}} with {{Multilayered Multielement Metagratings}}},''
  \emph{\BIBforeignlanguage{en}{IEEE Transactions on Antennas and
  Propagation}}, pp. 1--1, 2019.

\bibitem{ColeRefractionefficiencyHuygens2018}
M.~A. Cole, A.~Lamprianidis, I.~V. Shadrivov, and D.~A. Powell,
  ``\BIBforeignlanguage{en}{Refraction efficiency of {{Huygens}}' and
  bianisotropic terahertz metasurfaces},''
  \emph{\BIBforeignlanguage{en}{arXiv:1812.04725 [physics]}}, Dec. 2018.

\bibitem{fischer_causes_2013}
B.~E. Fischer, I.~J. {LaHaie}, M.~D. Huang, M.~H. A.~J. Herben, A.~C.~F.
  Reniers, and P.~F.~M. Smulders, ``Causes of discrepancies between
  measurements and em simulations of millimeter-wave antennas [measurements
  corner],'' \emph{{IEEE} Antennas and Propagation Magazine}, vol.~55, no.~6,
  pp. 139--149, 2013.

\bibitem{zelenchuk_millimeter-wave_2012}
D.~E. Zelenchuk, V.~Fusco, G.~Goussetis, A.~Mendez, and D.~Linton,
  ``Millimeter-wave printed circuit board characterization using substrate
  integrated waveguide resonators,'' \emph{{IEEE} Transactions on Microwave
  Theory and Techniques}, vol.~60, no.~10, pp. 3300--3308, 2012.

\bibitem{usta_effects_2019}
E.~Usta and N.~T. Tokan, ``Effects of surface finish material on
  millimeter-wave antenna performance,'' \emph{{IEEE} Transactions on
  Components, Packaging and Manufacturing Technology}, vol.~9, no.~5, pp.
  815--821, 2019.

\bibitem{holmberg_surface_2018}
M.~Holmberg, D.~Dancila, A.~Rydberg, B.~Hjörvarsson, U.~Jansson, J.~J.
  Marattukalam, N.~Johansson, and J.~Andresson, ``On {Surface} {Losses} in
  {Direct} {Metal} {Laser} {Sintering} {Printed} {Millimeter} and
  {Submillimeter} {Waveguides},'' \emph{Springer}, vol.~39, no.~6, pp.
  535--545, Feb. 2018.

\bibitem{curranMethodologyCombinedModeling2010}
B.~Curran, I.~Ndip, S.~Guttowski, and H.~Reichl, ``\BIBforeignlanguage{en}{A
  {{Methodology}} for {{Combined Modeling}} of {{Skin}}, {{Proximity}},
  {{Edge}}, and {{Surface Roughness Effects}}},''
  \emph{\BIBforeignlanguage{en}{IEEE Transactions on Microwave Theory and
  Techniques}}, vol.~58, no.~9, pp. 2448--2455, Sep. 2010.

\bibitem{yong-hoon_kim_scattering_1999}
{Yong-Hoon Kim}, {Ki-Seok Yang}, and {Sung-Hyun Kim}, ``Scattering
  characteristics of surface roughness in frequency and incident angle
  dependent at millimeter-wave,'' in \emph{1999 {Asia} {Pacific} {Microwave}
  {Conference}. {APMC}'99. {Microwaves} {Enter} the 21st {Century}.
  {Conference} {Proceedings} ({Cat}. {No}.99TH8473)}, vol.~3, Nov. 1999, pp.
  789--792 vol.3.

\bibitem{goldPhysicalSurfaceRoughness2017}
G.~Gold and K.~Helmreich, ``A {{Physical Surface Roughness Model}} and {{Its
  Applications}},'' \emph{IEEE Transactions on Microwave Theory and
  Techniques}, vol.~65, no.~10, pp. 3720--3732, Oct. 2017.

\bibitem{LaroucheReconciliationgeneralizedrefraction2012}
S.~Larouche and D.~R. Smith, ``Reconciliation of generalized refraction with
  diffraction theory.'' \emph{Optics letters}, vol.~37, no.~12, pp. 2391--3,
  Jun. 2012.

\bibitem{chen_theory_2018}
\BIBentryALTinterwordspacing
M.~Chen, E.~Abdo-Sánchez, A.~Epstein, and G.~V. Eleftheriades, ``Theory,
  design, and experimental verification of a reflectionless bianisotropic
  huygens' metasurface for wide-angle refraction,'' \emph{Physical Review B},
  vol.~97, no.~12, p. 125433, 2018. [Online]. Available:
  \url{https://link.aps.org/doi/10.1103/PhysRevB.97.125433}
\BIBentrySTDinterwordspacing

\bibitem{lavigne_susceptibility_2018}
G.~Lavigne, K.~Achouri, V.~S. Asadchy, S.~A. Tretyakov, and C.~Caloz,
  ``Susceptibility {Derivation} and {Experimental} {Demonstration} of
  {Refracting} {Metasurfaces} {Without} {Spurious} {Diffraction},'' \emph{IEEE
  Transactions on Antennas and Propagation}, vol.~66, no.~3, pp. 1321--1330,
  Mar. 2018.

\bibitem{olkHighlyAccurateFullypolarimetric2017}
A.~Olk, K.~B. Khadhra, and T.~Spielmann, ``Highly accurate fully-polarimetric
  radar cross section facility for mono- and bistatic measurements at
  {{W}}-band frequencies,'' in \emph{Antenna {{Measurement Techniques
  Association Symposium}} ({{AMTA}})}, Oct. 2017, pp. 1--6, iEEE Proceedings.

\bibitem{gopalakrishnanStudyEffectSurface2016}
S.~Gopalakrishnan, A.~Dasgupta, and D.~R. Nair, ``\BIBforeignlanguage{en}{Study
  of the {{Effect}} of {{Surface Roughness}} on the {{Performance}} of {{RF
  MEMS Capacitive Switches Through}} 3-{{D Geometric Modeling}}},''
  \emph{\BIBforeignlanguage{en}{IEEE Journal of the Electron Devices Society}},
  vol.~4, no.~6, pp. 451--458, Nov. 2016.

\bibitem{cst}
{CST Microwave Studio}, ``Darmstadt - {Germany},'' 2018.

\bibitem{capolino_equivalent_2013}
\BIBentryALTinterwordspacing
F.~Capolino, A.~Vallecchi, and M.~Albani, ``\BIBforeignlanguage{en}{Equivalent
  {Transmission} {Line} {Model} {With} a {Lumped} {X}-{Circuit} for a
  {Metalayer} {Made} of {Pairs} of {Planar} {Conductors}},''
  \emph{\BIBforeignlanguage{en}{IEEE Transactions on Antennas and
  Propagation}}, vol.~61, no.~2, pp. 852--861, Feb. 2013. [Online]. Available:
  \url{http://ieeexplore.ieee.org/document/6331512/}
\BIBentrySTDinterwordspacing

\bibitem{yu_light_2011}
\BIBentryALTinterwordspacing
N.~Yu, P.~Genevet, M.~A. Kats, F.~Aieta, J.-P. Tetienne, F.~Capasso, and
  Z.~Gaburro, ``\BIBforeignlanguage{en}{Light {Propagation} with {Phase}
  {Discontinuities}: {Generalized} {Laws} of {Reflection} and {Refraction}},''
  \emph{\BIBforeignlanguage{en}{Science}}, vol. 334, no. 6054, pp. 333--337,
  Oct. 2011. [Online]. Available:
  \url{http://science.sciencemag.org/content/334/6054/333}
\BIBentrySTDinterwordspacing

\bibitem{hiebel_fundamentals_2011}
M.~Hiebel, \emph{Fundamentals of {Vector} {Network} {Analysis}}, 5th~ed.\hskip
  1em plus 0.5em minus 0.4em\relax Munich, Germany: Rohde \& Schwarz, 2011.

\end{thebibliography}

%


\begin{IEEEbiography}[{\includegraphics[width=1in,height=1.25in,clip,keepaspectratio]{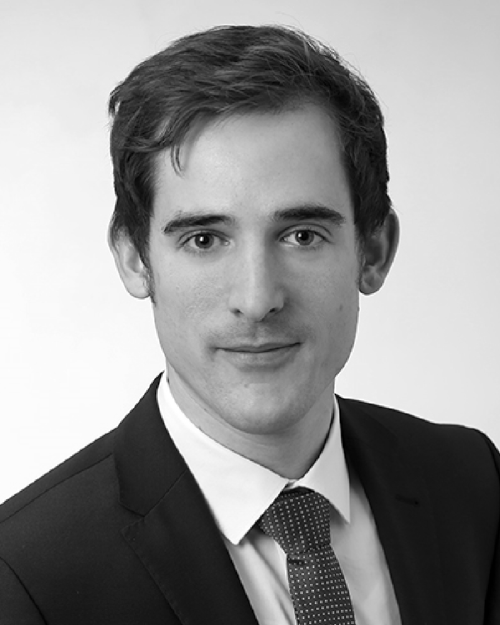}}]{Andreas E. Olk}
	 received the B.Sc and the M.Sc. degree in physics from RWTH University, Aachen, Germany, in 2012 and 2014, respectively. Since 2015, he is affiliated with IEE S.A., Luxembourg. Since 2018, he is pursuing a Ph.D. in electrical engineering with the University of New South Wales, Canberra, Australia.  His current research interests include metamaterials, millimeter-wave technology, radar and automotive sensing.
\end{IEEEbiography}

\begin{IEEEbiography}[{\includegraphics[width=1in,height=1.25in,clip,keepaspectratio]{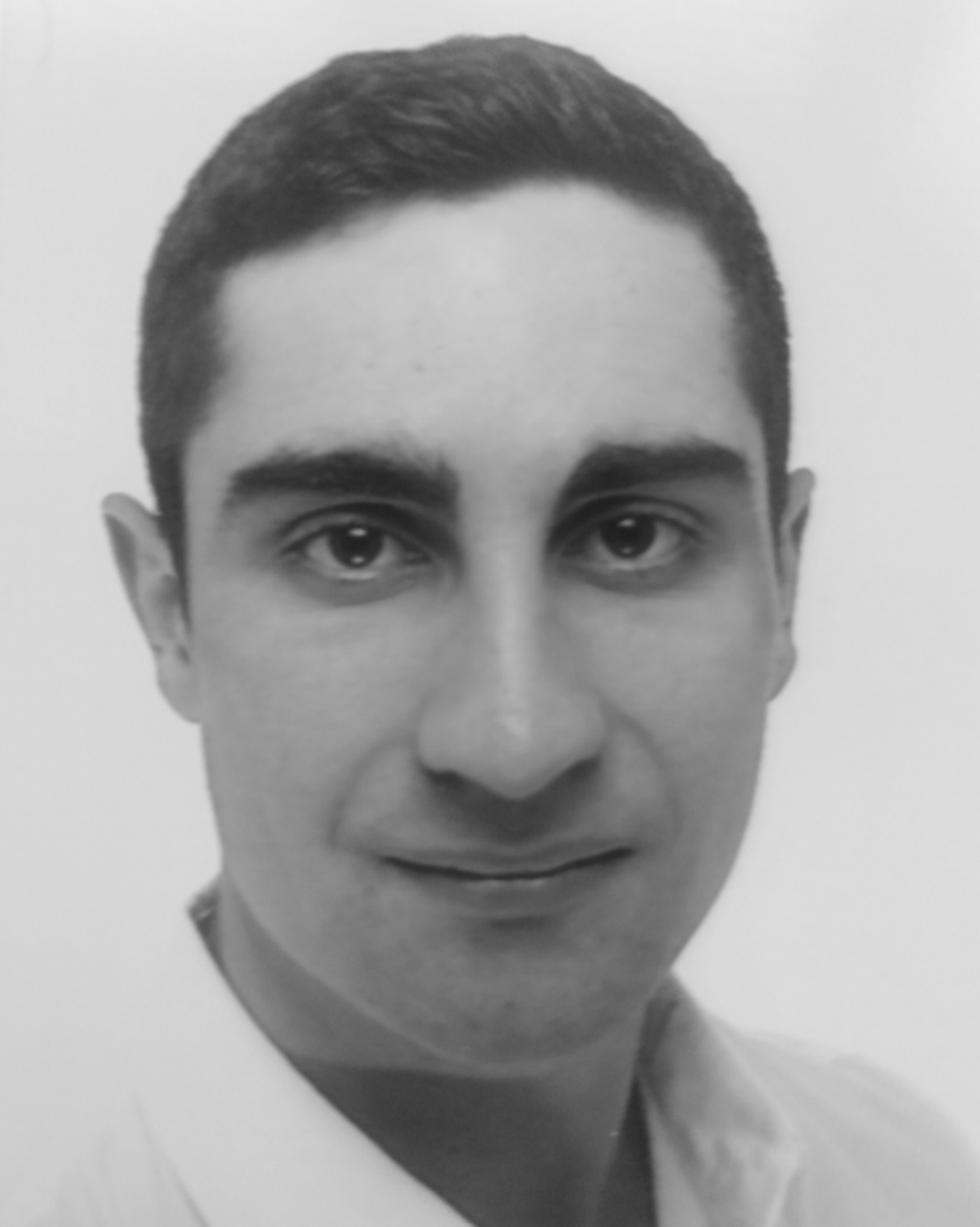}}]{Pierre~E.~M.~Macchi}
	received the B.Sc. in physics and electronics from Grenoble INP Phelma University, Grenoble, France in 2018. He is currently pursuing the M.Sc. degree IPHY, Physical engineering for photonics and microelectronics from Grenoble INP Phelma University, Grenoble, France.	
\end{IEEEbiography}

\begin{IEEEbiography}[{\includegraphics[width=1in,height=1.25in,clip,keepaspectratio]{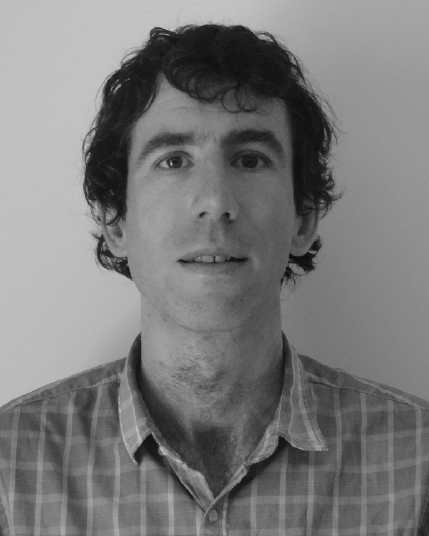}}]{David A. Powell} (M’05-SM'17) is a Senior Lecturer at the School of Engineering and Information Technology, University of New South Wales, Canberra, Australia.
    His research on metamaterials has covered the microwave, millimeter-wave, terahertz, and near-infrared wavelength ranges, in addition to work on acoustic metamaterials.	He received the Bachelor of Computer Science and Engineering from Monash University, Melbourne, Australia, in 2001, and the Ph.D. degree in Electronic and Communications Engineering from RMIT University, Melbourne, Australia in 2006. Between 2006 and 2017, he was a researcher with the Nonlinear Physics Centre at the Australian National University. 
\end{IEEEbiography}




\end{document}